\documentclass{osa-article}

\journal{oe}

\articletype{Research Article}

\begin{document}

\title{Spectrometric detection of weak forces in cavity optomechanics}

\author{Yue-Hui Zhou,\authormark{1} Qing-Shou Tan,\authormark{2} Xi-Ming Fang,\authormark{1} Jin-Feng Huang,\authormark{1,*} and Jie-Qiao Liao\authormark{1,$\dagger$}}

\address{\authormark{1}Key Laboratory of Low-Dimensional Quantum Structures and Quantum Control of
Ministry of Education, Key Laboratory for Matter Microstructure and Function of Hunan Province, Department of Physics and Synergetic Innovation
Center for Quantum Effects and Applications, Hunan Normal University,
Changsha 410081, China\\
\authormark{2}College of Physics and Electronic Engineering, Hainan Normal University, Haikou 571158, China}

\email{\authormark{*}jfhuang@hunnu.edu.cn.} 
\email{\authormark{$\dagger$}jqliao@hunnu.edu.cn} 

\begin{abstract}
We propose a spectrometric method to detect a classical weak force acting upon the moving end mirror in a cavity optomechanical system. The force changes the equilibrium position of the end mirror, and thus the resonance frequency of the cavity field depends on the force to be detected. As a result, the magnitude of the force can be inferred by analyzing the single-photon emission and scattering spectra of the optomechanical cavity. Since the emission and scattering processes are much faster than the characteristic mechanical dissipation, the influence of the mechanical thermal noise is negligible in this spectrometric detection scheme. We also extent this spectrometric method to detect a monochromatic oscillating force by utilizing an optomechanical coupling modulated at the same frequency as the force.
\end{abstract}

\section{Introduction}

Cavity optomechanical systems provide a superior platform for study of the radiation-pressure interaction between electromagnetic fields and mechanical oscillation~\cite{Kippenberg2008Science,Aspelmeyer2012PhysTod,Aspelmeyer2014RMP,Bowen2016book}. How to understand, manipulate, and exploit the optomechanical (radiation-pressure) interaction is an important research topic in this field. Until now, great theoretical and experimental advances have been made in various optomechanical topics such as ground-state cooling of moving mirrors~\cite{Wilson-Rae2007PRL,Marquardt2007PRL,Genes2008PRA,Teufel2011Nature,Chan2011Nature,Lai2018PRA}, optomechanically induced transparency~\cite{Agarwal2010PRA,Weis2010Science,Safavinaeini2011Nature}, optomechanical entanglement~\cite{Bose1997PRA,Vitali2007PRL,Hartmann2008PRL,Tian2013PRL,Wang2013PRL,Palomaki2013Science}, and other applications based on optomechanical technology~\cite{Metcalfe2014APR}.

From the view point of applications, cavity optomechanical systems are elegant candidates for implementation of ultrasensitive detection of weak forces and tiny displacement~\cite{Caves1980RMP,Bocko1996RMP,Teufel2009NN,Schliesser2009NatPhys,Gavartin2012NN}, ultrasensitive accelerometer~\cite{Krause2012NatPhot}, magnetometer~\cite{Forstner2012PRL,Li2018Opt}, and other precision measurement schemes. This is because the two elements of cavity optomechanical systems, the optical field and the mechanical oscillator, are excellent systems for realization of the signal detection and signal sensing~\cite{Blencowe2004PhysRep,Poot2012PhysRep}. For example, the optical degree of freedom can be used to implement ultrasensitive phase measurement in various interferometers~\cite{Caves1980RMP}, and the mechanical resonators are widely used for sensing weak forces~\cite{Braginsky1977book,Braginsky1992book}. Owing to these features, the optomechanical systems have been suggested to study high-sensitivity interferometric measurements in quantum metrology. Recently, several schemes have been proposed to implement the detection of weak forces with the optomechanical systems working in the strong-driving and linearization regime~\cite{Vitali2001PRA,Fermani2004PRA,Lucamarini2006PRA,Ma2015Srep,Motazedifard2016NJP,Zhang2017NJP,Armata2017PRA,Li2018PRA}. In usual force detection schemes, the detections are performed based on the steady state of the system, and hence the thermal noise of the mechanical mode will affect the detection precision.

Currently, several optomechanical systems are approaching the single-photon strong-coupling regime~\cite{Gupta2007PRL,Brennecke2008Science,Eichenfield2009Nature,Pirkkalaine2015NatComm}, and much attention has been paid to the study of optomechanical interaction at the single-photon level~\cite{Rabl2011PRL,Nunnenkamp2011PRL,Liao2012PRA,Qian2012PRL,Ludwig2012PRL,Stannigel2012PRL,Kronwald2013PRA,Liao2013PRA,Liao2014PRA,Lu2015PRL}. In this regime, the radiation pressure induced by a single photon could lead to observable physical effects. Therefore, a natural question is if we can detect a weak force by utilizing the sensitivity of a single photon. In this paper, we propose a reliable method to detect weak forces by analyzing the spectrum of a single photon either emitted or scattered by the optomechanical cavity. The external force acting upon the moving end mirror leads to a change of the cavity resonance frequency, which could be inferred from the frequency shift of the single-photon emission and scattering spectra. This physical mechanism motivates us to study the spectrometric detection of weak external forces. In typical optomechanical systems, the decay rate of the moving mirror is much smaller than that of the cavity field, and hence the influence of the thermal noise from the mechanical resonator can be neglected in the single-photon emission and scattering processes~\cite{Bowen2016book,Liao2012PRA,Liao2013PRA,Liao2014PRA}. This feature is an advantage of the spectrometric method over other schemes based on linearized optomechanical interaction and steady-state measurement. Note that the single-photon emission and scattering spectra in an optomechanical cavity have been suggested to evaluate the optomechanical coupling strength~\cite{Liao2012PRA} and to implement quantum state reconstruction of the mechanical mode~\cite{Liao2014PRA}.

The rest of this paper is organized as follows. In Sec.~\ref{sec2}, we introduce the optomechanical model and present the Hamiltonian. In Sec.~\ref{sec3}, we obtain the exact solutions of single-photon emission and scattering under the Wigner-Weisskopf framework, and analyze how to infer the force magnitude from the shift of the single-photon spectra. In Sec.~\ref{sec4}, we extend the spectrometric method to detect a monochromatic oscillating force. Finally, we conclude this work with some discussions in Sec.~\ref{sec5}.

\section{Model and Hamiltonian \label{sec2}}

\begin{figure}[tbp]
\center
\includegraphics[width=7.4cm]{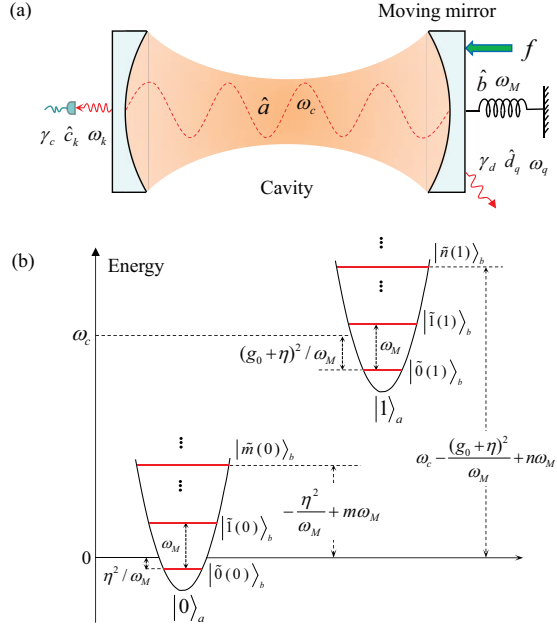}
\caption{(a) Schematic of the Fabry-P\'{e}rot-type optomechanical cavity formed by a fixed end mirror and a moving end mirror. The classical constant force $f$ acting on the moving mirror is detected by analyzing the single-photon emission and scattering spectra. The outside fields corresponding to the detected photons and the leaked photons are described by two independent vacuum baths $\hat{c}_{k}$ and $\hat{d}_{q}$, respectively. (b) The eigensystem of the optomechanical cavity limited within the zero- and one-photon subspaces.}
\label{Fig1}
\end{figure}

We start with introducing the Fabry-P\'{e}rot-type optomechanical cavity [see Fig.~\ref{Fig1}(a)], which is formed by a fixed end mirror and a moving end mirror. Here we consider the adiabatic regime where the mechanical resonance frequency is much smaller than the free spectral range of the cavity, then there are no photons scattering among these cavity modes, and hence one can only focus on a single field mode in the cavity~\cite{Law1994PRA,Law1995PRA}. We also assume that the oscillation amplitude of the moving mirror is much smaller than the rest length of the cavity, then one can perform a linear expansion of the cavity frequency as a function of the oscillation displacement. Therefore, the Hamiltonian of the optomechanical cavity under the action of the constant force reads ($\hbar =1$)
\begin{equation}
\hat{H}_{\text{opc}}=\omega_{c}\hat{a}^{\dagger}\hat{a}+\omega_{M}\hat{b}^{\dagger}\hat{b}-(g_{0}\hat{a}^{\dagger}\hat{a}+\eta)(\hat{b}^{\dagger}+\hat{b}),
\end{equation}
where $\hat{a}$ ($\hat{a}^{\dagger}$) and $\hat{b}$ ($\hat{b}^{\dagger}$) are, respectively, the annihilation (creation) operators of the cavity field and the mechanical mode, with the corresponding resonance frequencies $\omega_{c}$ and $\omega _{M}$, the $g_{0}$ term describes the radiation-pressure coupling between the single-mode cavity field and the mechanical oscillation, with $g_{0}$ being the single-photon optomechanical-coupling strength. The term $-\eta(\hat{b}^{\dagger}+\hat{b})$ describes a classical constant force $f$ acting upon the moving end mirror, where the coupling strength is defined by $\eta=fx_{0}$, with $x_{0}=1/\sqrt{2M\omega _{M}}$ being the zero-point fluctuation of the moving mirror with mass $M$.

To understand the physical mechanism of the force detection, we first analyze the eigensystem of the Hamiltonian $\hat{H}_{\text{opc}}$. Let us denote $|m\rangle_{a}$ and $|j\rangle_{b}$ ($m,j=0,1,2,\cdots$) as number states of the cavity field and the mechanical resonator, respectively, then the eigen-equation of the Hamiltonian $\hat{H}_{\text{opc}}$ can be obtained as
\begin{equation}
\hat{H}_{\text{opc}}|m\rangle_{a}|\tilde{j}(m)\rangle_{b}=E_{m,j}|m\rangle_{a}|\tilde{j}(m)\rangle_{b},\label{eigeneq}
\end{equation}
where $E_{m,j}=m\omega _{c}+j\omega _{M}-(g_{0}m+\eta )^{2}/\omega _{M}$ are the eigenvalues and $|\tilde{j}(m)\rangle_{b}=\hat{D}_{b}(\beta_{m})|j\rangle_{b}=\exp[\beta_{m}(\hat{b}^{\dagger}-\hat{b})]|j\rangle_{b}$ are $m$-photon-dependent displaced phonon number states with $\beta_{m}=(g_{0}m+\eta)/\omega_{M}$. For studying the single-photon spectra, in Fig.~\ref{Fig1}(b) we show the eigensystem of the optomechanical cavity in the zero- and single-photon state subspaces. Here we can see that, corresponding to the cavity field states $|0\rangle_{a}$ and $|1\rangle_{a}$, the ground-state energies of the mechanical degree of freedom move down by the values $\eta^{2}/\omega_{M}$ and $(g_{0}+\eta)^{2}/\omega_{M}$, respectively.

The relationship between the external force and the energy shifts motivates us to detect the force by analyzing the spectra of a single photon emitted and scattered by the optomechanical cavity. To calculate the single-photon spectra, the detected continuous fields outside the cavity should be treated quantum mechanically. However, there always exist some photons leaking through other undetected channels in a realistic optomechanical system. To include this photon loss effect, we introduce another vacuum bath to the cavity field so that the photons leaking through other channels can be covered. Based on the above analyses, the Hamiltonian of the two vacuum baths can be written as $\hat{H}_{\text{B}}=\sum_{k}\omega _{k}\hat{c}_{k}^{\dagger }\hat{c}_{k}+\sum_{q}\omega _{q}\hat{d}_{q}^{\dagger }\hat{d}_{q}$ where $\hat{c}_{k}$ ($\hat{d}_{q}$) and $\hat{c}_{k}^{\dagger}$ ($\hat{d}^{\dag}_{q}$) are the annihilation and creation operators of the $k$th ($q$th) mode of the detected outside fields (undetected vacuum bath) with the resonance frequency $\omega_{k}$ ($\omega_{q}$). The two vacuum baths are independent of each other and are coupled to the cavity through the photon-hopping interaction. The interaction Hamiltonian describing the couplings between the optomechanical cavity and the two vacuum baths can be written as
\begin{equation}
\hat{H}_{\text{int}}=\sum_{k}\xi_{k}(\hat{a}^{\dagger}\hat{c}_{k}+\hat{c}_{k}^{\dagger}\hat{a})
+\sum_{q}\chi_{q}(\hat{a}^{\dagger}\hat{d}_{q}+\hat{d}_{q}^{\dagger}\hat{a}),
\end{equation}
where $\xi_{k}=\xi(\omega_{k})$ and $\chi_{q}=\chi(\omega_{q})$ are the strengths of the photon hopping corresponding to the detected and undetected vacuum baths, respectively. Based on the above analysis, the total Hamiltonian can be written as
\begin{equation}
\hat{H}=\hat{H}_{\text{opc}}+\hat{H}_{\text{B}}+\hat{H}_{\text{int}},\label{Htotorig}
\end{equation}
which describes the whole system including the optomechanical cavity and the two vacuum baths. Note that the mechanical thermal noise is neglected in our scheme because the influence of the mechanical thermal noise is negligible when the single-photon emission and scattering are completed. This treatment is valid when the mechanical reservoir is at a low temperature and the mechanical damping is slow compared to all other time scales~\cite{Bowen2016book,Liao2012PRA,Liao2013PRA,Liao2014PRA}.

\section{Spectrometric detection of the force \label{sec3}}

In this section, we study how to detect the external force acting on the moving mirror by analyzing the spectrum of a single photon either emitted or scattered by the optomechanical cavity.

\subsection{Equations of motion}

In a rotating frame with respect to $\hat{H}_{0}=\omega_{c}(\hat{a}^{\dagger}\hat{a}+\sum_{k}\hat{c}_{k}^{\dagger}\hat{c}_{k}+\sum_{q}\hat{d}_{q}^{\dagger}\hat{d}_{q})$, Hamiltonian~(\ref{Htotorig}) becomes
\begin{eqnarray}
\hat{H}_{I} &=&\omega _{M}\hat{b}^{\dagger }\hat{b}-g_{0}\hat{a}^{\dagger }%
\hat{a}(\hat{b}^{\dagger }+\hat{b})-\eta (\hat{b}^{\dagger }+\hat{b}%
)+\sum_{k}\Delta _{k}\hat{c}_{k}^{\dagger }\hat{c}_{k}+\sum_{q}\Delta _{q}%
\hat{d}_{q}^{\dagger }\hat{d}_{q}  \nonumber \\
&&+\sum_{k}\xi _{k}(\hat{c}_{k}^{\dagger }\hat{a}+\hat{a}^{\dagger }\hat{c}%
_{k})+\sum_{q}\chi _{q}(\hat{a}^{\dagger }\hat{d}_{q}+\hat{d}_{q}^{\dagger }%
\hat{a}),  \label{eq1iniH}
\end{eqnarray}
where $\Delta_{k}=\omega_{k}-\omega_{c}$ and $\Delta_{q}=\omega_{q}-\omega_{c}$. In this system, the total photon number operator, defined by $\hat{N}=\hat{a}^{\dagger}\hat{a}+\sum_{k}\hat{c}_{k}^{\dagger}\hat{c}_{k}+\sum_{q}\hat{d}_{q}^{\dagger}\hat{d}_{q}$, is a conserved quantity because of $[\hat{N},\hat{H}]=0$. Therefore, we restrict the system within the single-photon subspace, in which a general pure state of the total system can be expressed as
\begin{eqnarray}
|\varphi (t)\rangle  &=&\sum_{n=0}^{\infty }A_{n}(t)|1\rangle _{a}|\tilde{n}%
(1)\rangle _{b}|\emptyset \rangle _{c}|\emptyset \rangle
_{d}+\sum_{n=0}^{\infty }\sum_{k}B_{n,k}(t)|0\rangle _{a}|\tilde{n}%
(0)\rangle _{b}|1_{k}\rangle _{c}|\emptyset \rangle _{d} \notag \\
&&+\sum_{n=0}^{\infty }\sum_{q}C_{n,q}(t)|0\rangle _{a}|\tilde{n}(0)\rangle
_{b}|\emptyset \rangle _{c}|1_{q}\rangle _{d},  \label{state}
\end{eqnarray}
where $|\emptyset\rangle_{c}$ and $|\emptyset\rangle_{d}$ are the vacuum states of the two baths, $|1_{k}\rangle_{c}=\hat{c}_{k}^{\dagger}|\emptyset\rangle_{c}$ and $|1_{q}\rangle_{d}=\hat{d}_{q}^{\dagger}|\emptyset\rangle_{d}$ are the single-photon basis states of the two baths, $A_{n}(t)$, $B_{n,k}(t)$, and $C_{n,q}(t)$ are the corresponding probability amplitudes. Inserting Hamiltonian~(\ref{eq1iniH}) and single-photon state~(\ref{state}) into the Schr\"{o}dinger equation, we obtain the following equations of motion for these probability amplitudes
\begin{subequations}
\label{eqofmotion}
\begin{align}
\dot{A}_{m}(t)=& -i[m\omega _{M}-(g_{0}+\eta )^{2}/\omega _{M}]A_{m}(t)
\notag \\
& -i\sum_{n=0}^{\infty }\,_{b}\langle \tilde{m}(1)|\tilde{n}(0)\rangle _{b}%
\left[ \sum_{k}\xi _{k}B_{n,k}(t)+\sum_{q}\chi _{q}C_{n,q}(t)\right] , \\
\dot{B}_{m,k}(t)=& -i(m\omega _{M}+\Delta _{k}-\eta ^{2}/\omega
_{M})B_{m,k}(t)-i\xi _{k}\sum_{n=0}^{\infty }\,_{b}\langle \tilde{m}(0)|%
\tilde{n}(1)\rangle _{b}A_{n}(t), \\
\dot{C}_{m,q}(t)=& -i(m\omega _{M}+\Delta _{q}-\eta ^{2}/\omega
_{M})C_{m,q}(t)-i\chi _{q}\sum_{n=0}^{\infty }\,_{b}\langle \tilde{m}(0)|%
\tilde{n}(1)\rangle _{b}A_{n}(t).
\end{align}
\end{subequations}
Here these transition matrix elements are determined by the coupling strengths $\xi_{k}$, $\chi_{q}$, and the Franck-Condon factors $_{b}\langle \tilde{m}(1)|\tilde{n}(0)\rangle_{b}=\;_{b}\langle \tilde{n}(0)|\tilde{m}(1)\rangle_{b}^{*}$ and $_{b}\langle\tilde{m}(0)|\tilde{n}(1)\rangle_{b}=\;_{b}\langle \tilde{n}(1)|\tilde{m}(0)\rangle^{*}_{b}$, which are the overlap between the zero-photon displaced phonon number states and the single-photon displaced phonon number states.
The value of these Franck-Condon factors can be calculated by~\cite{Oliveira1990PRA,Crisp1992PRA}
\begin{equation}
_{b}\!\langle \tilde{m}(0)|\tilde{n}(1)\rangle _{b}=\left\{
\begin{array}{c}
\sqrt{\frac{m!}{n!}}e^{-\frac{\beta ^{2}}{2}}(-\beta
)^{n-m}L_{m}^{n-m}(\beta ^{2}),\hspace{0.05cm}n\geq m, \\
\sqrt{\frac{n!}{m!}}e^{-\frac{\beta ^{2}}{2}}\beta ^{m-n}L_{n}^{m-n}(\beta
^{2}),\hspace{0.3cm}m>n,%
\end{array}%
\right.   \label{matrixelemD}
\end{equation}
where $\beta\equiv\beta_{1}-\beta_{0}=g_{0}/\omega_{M}$.

In principle, we can solve Eq.~(\ref{eqofmotion}) for these probability amplitudes under given initial conditions.
Corresponding to the single-photon emission and scattering cases, the single photon is assumed to be initially in the cavity and the detected continuous fields outside the cavity, respectively. Based on these probability amplitudes, the steady state of the system can be obtained by taking the long-time limit of state~(\ref{state}), and we can calculate the single-photon spectra by evaluating the photon number distribution in the detected continuous fields.
In the next two subsections, we will consider the force detection based on single-photon emission and scattering spectra, respectively.

\subsection{Force detection by emission spectrum \label{secabpn}}

In the single-photon emission case, the single photon is initially in the cavity. For the moving mirror, its state could be an arbitrary state. In our discussions, we will consider the number state, coherent state, and thermal state. Below, we will first solve Eq.~(\ref{eqofmotion}) corresponding to an initial number state $|m_{0}\rangle_{b}$ of the moving mirror. Once the solution for this number-state case is found, the solution for a general initial state of the moving mirror can be obtained accordingly by superposition. Corresponding to the initial state of the total system $|\varphi (0)\rangle =|1\rangle _{a}|m_{0}\rangle _{b}|\emptyset \rangle _{c}|\emptyset \rangle _{d}$, the initial condition of these probability amplitudes is given by $A_{m}(0)=\,_{b}\langle \tilde{m}(1)|m_{0}\rangle _{b}$ and $B_{m,k}(0)=C_{m,q}(0)=0$.
Based on the initial condition, the transient solution of these probability amplitudes can be obtained. For studying single-photon spectra, here we only show the long-time solution
\begin{subequations}
\label{SOLlongt}
\begin{align}
A_{m_{0},m}(t\rightarrow \infty )=& 0, \\
B_{m_{0},m,k}(t\rightarrow \infty )=& \sqrt{\frac{\gamma _{c}}{2\pi \mu
(\omega _{k})}}\sum_{n=0}^{\infty }\frac{\,_{b}\langle \tilde{m}(0)|\tilde{n}%
(1)\rangle _{b}\,_{b}\langle \tilde{n}(1)|m_{0}\rangle _{b}}{\Delta
_{k}+\lambda -(n-m)\omega _{M}+i\frac{\gamma }{2}}e^{-i(\Delta _{k}+m\omega
_{M}-\zeta )t}, \\
C_{m_{0},m,q}(t\rightarrow \infty )=& \sqrt{\frac{\gamma _{d}}{2\pi \nu
(\omega _{q})}}\sum_{n=0}^{\infty }\frac{\,_{b}\langle \tilde{m}(0)|\tilde{n}%
(1)\rangle _{b}\,_{b}\langle \tilde{n}(1)|m_{0}\rangle _{b}}{\Delta
_{q}+\lambda -(n-m)\omega _{M}+i\frac{\gamma }{2}}e^{-i(\Delta _{q}+m\omega
_{M}-\zeta )t}.
\end{align}
\end{subequations}
Here $\gamma=\gamma_{c}+\gamma_{d}$ is the total decay rate of the cavity field, $\gamma_{c}=2\pi\mu(\omega_{c})\xi^{2}(\omega_{c})$ and $\gamma_{d}=2\pi\nu(\omega_{c})\chi^{2}(\omega_{c})$ are, respectively, the decay rates corresponding to the detected and undetected continuous fields, with $\mu(\omega_{k})$ and $\nu(\omega_{q})$ being the densities of state corresponding to the two decay channels. In this work, the cavity frequency is much larger than the mechanical frequency and we treat the emission of the cavity field under the Wigner-Weisskopf framework. Namely, we assume that the cavity bath spectral density is flat over the whole range of relevant phonon sidebands, then the cavity dissipation can be described by two rates $\gamma_{c}$ and $\gamma_{d}$. In particular, in all the calculations relating the integral of variables $\omega_{k}$ and $\omega_{q}$, we can make the replacements $\omega_{k}\rightarrow\omega_{c}$ and $\omega_{q}\rightarrow\omega_{c}$ in the variables $\mu(\omega_{k})$,  $\xi(\omega_{k})$, $\nu(\omega_{q})$,  and $\chi(\omega_{q})$. The parameter $\zeta=\eta^{2}/\omega_{M}$ is the ground-state energy shift of the mirror induced by the force (i.e., the energy shift corresponding to the zero-photon state), and $\lambda=(g_{0}^{2}+2g_{0}\eta)/\omega_{M}=(g_{0}+\eta)^{2}/\omega_{M}-\eta^{2}/\omega_{M}$ is the difference between the two ground-state energy shifts of the mirror corresponding to the zero-photon and single-photon states, as shown in Fig.~\ref{Fig1}(b). Note that here we add an index $m_{0}$ to these probability amplitudes in Eq.~(\ref{SOLlongt}) for marking the corresponding initial state $|m_{0}\rangle_{b}$. It should be pointed out that the long-time limit actually refers to the time scale of $1/\gamma\ll t\ll1/\gamma_{M}$ with $\gamma_{M}$ being the decay rate of the moving mirror. This is because $\gamma\gg\gamma_{M}$ in most realistic optomechanical systems, and thus the damping of the mechanical motion is negligible until the single photon has been completely leaked out of the cavity.

In the long-time limit, the single photon leaks completely out of the cavity, then $A_{m_{0},m}(\infty)=0$. Since the single photon is initially in the cavity, we can expand the initial state of the moving mirror with the single-photon displaced number states as $|\varphi(0)\rangle=\sum_{n=0}^{\infty}\,_{b}\langle\tilde{n}(1)|m_{0}\rangle_{b}|1\rangle_{a}|\tilde{n}(1)\rangle_{b}|\emptyset\rangle_{c}|\emptyset\rangle_{d}$.
When the photon is emitted out of the cavity, the cavity field and the moving mirror experience the transitions $|1\rangle_{a}\rightarrow|0\rangle_{a}$ and $|\tilde{n}(1)\rangle_{b}\rightarrow|\tilde{m}(0)\rangle_{b}$, respectively. This transition process can be seen from the numerator of the probability amplitudes $B_{m_{0},m,k}(t\rightarrow\infty)$ and $C_{m_{0},m,q}(t\rightarrow\infty)$. In the single-photon emission process, the phonon sidebands will participate the transition process, and thus the frequency of the emitted photon will be determined by the resonance conditions
\begin{equation}
\Delta_{k(q)}=(n-m)\omega _{M}-\frac{1}{\omega _{M}}(g_{0}^{2}+2g_{0}\eta)  \label{resfreqcyDeltak}.
\end{equation}
These two resonance conditions can also be obtained from the real part of the pole of the denominator in $B_{m_{0},m,k}(t\rightarrow\infty) $ and $C_{m_{0},m,q}(t\rightarrow\infty)$.

In this system, the single-photon spectra can be obtained by calculating the final reservoir occupation distribution~\cite{Linington2008PRA}, namely the probability density for finding the single photon in the $k$th mode of these detected continuous fields outside the cavity. When the moving mirror is initially in either a pure state $\sum_{m_{0}=0}^{\infty}R_{m_{0}}(0)|m_{0}\rangle_{b}$ or a mixed state $\sum_{m_{0}=0}^{\infty}P_{m_{0}}(0)|m_{0}\rangle_{b}\!\,_{b}\langle m_{0}|$, the corresponding single-photon emission spectrum takes the form as~\cite{Liao2012PRA}
\begin{subequations}
\label{spectraform}
\begin{align}
S(\Delta_{k})&=\mu(\omega_{k})\sum_{m=0}^{\infty}\left\vert\sum_{m_{0}=0}^{\infty}R_{m_{0}}(0)B_{m_{0},m,k}(t\rightarrow\infty)\right \vert^{2},\label{purespectra}\\
S(\Delta_{k})&=\mu(\omega_{k})\sum_{m_{0}=0}^{\infty}P_{m_{0}}\sum_{m=0}^{\infty}\left \vert B_{m_{0},m,k}(t\rightarrow\infty)\right\vert^{2}.\label{mixedspectra}
\end{align}
\end{subequations}

\begin{figure}[tbp]
\center
\includegraphics[width=7.8cm]{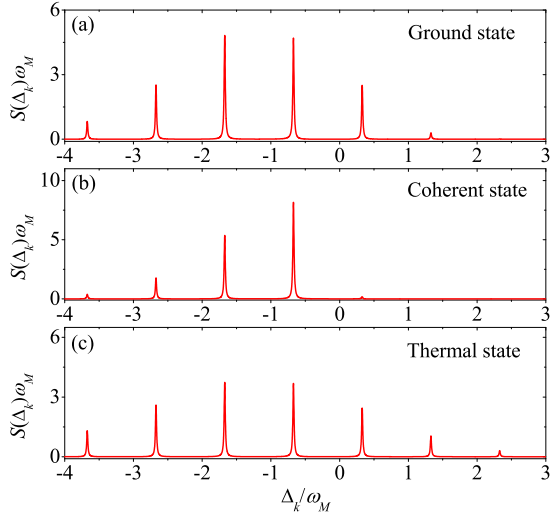}
\caption{The scaled single-photon emission spectrum $S(\Delta_{k})\omega_{M}$ as a function of $\Delta_{k}/\omega_{M}$ when the moving mirror is initially in different states: (a) ground state $|0\rangle_{b}$, (b) coherent state $|\alpha\rangle_{b}$ with $\alpha=1$, and (c) thermal state $\rho_{b}^{\text{th}}$ with average thermal phonon number $\bar{n}_{\text{th}}=1$. The used parameters are $g_{0}/\omega_{M}=0.8$, $\gamma_{c}/\omega_{M}=\gamma_{d}/\omega_{M}=0.01$, and $\eta/\omega_{M}=0.02$.}
\label{Fig2}
\end{figure}

The resonance condition in Eq.~(\ref{resfreqcyDeltak}) shows that these resonance peaks of the detected continuous fields are located at $\Delta_{k}=(n-m)\omega_{M}-(g_{0}^{2}+2g_{0}\eta)/\omega_{M}$. The external force $f$ causes a frequency shift $-2g_{0}\eta/\omega_{M}=-2g_{0}fx_{0}/\omega_{M}$ of the resonance peaks in the spectrum. The magnitude of the force can be inferred from the frequency shift of these resonance sidebands. In particular, when $m=n$, there always exists a peak at $\Delta_{k}=-(g_{0}^{2}+2g_{0}\eta)/\omega_{M}$. This peak is considered as the zero-phonon line (ZPL)~\cite{Rabl2011PRL} because it corresponds to the no-phonon-assisted transition.

In this force detection scheme, we expect that the system works in the deep-resolved-sideband regime $\omega_{M}\gg\gamma$ and the single-photon strong-coupling regime $g_{0} \gg\gamma$ so that these sideband peaks are sufficient sharp to be resolved and two most neighboring peaks are separated from each other very clearly~\cite{Rabl2011PRL,Nunnenkamp2011PRL,Liao2012PRA}. In Fig.~\ref{Fig2}, we plot the single-photon emission spectrum as a function of $\Delta_{k}$ when the moving mirror is initially in various states: ground state $|0\rangle_{b}$, coherent state $|\alpha\rangle_{b}=e^{-\frac{|\alpha|^{2}}{2}}\sum_{n=0}^{\infty}\frac{\alpha^{n}}{\sqrt{n!}}|n\rangle_{b}$ with $\alpha=1$, and thermal state $\rho _{b}^{\text{th}}=\sum_{n=0}^{\infty }\frac{\bar{n}_{\text{th}}^{n}}{(\bar{n}_{\text{th}}+1)^{n+1}}|n\rangle _{b}\,_{b}\langle n|$ with average thermal phonon number $\bar{n}_{\text{th}}=1$. Here we see some phonon sideband peaks in the spectrum. The location of these sideband peaks is independent of the initial states of the moving mirror, but the height of these peaks depends on the mechanical states. This is because the expansion coefficients of the superposition components $|\tilde{n}(1)\rangle_{b}$ are different for different initial states of the mirror. In addition, we see from Eqs.~(\ref{SOLlongt}) and (\ref{spectraform}) that, for a given $\gamma$, the magnitude of the decay rate $\gamma_{c}$ affects the height of the spectrum because the numerator of the emission spectrum is proportional to $\gamma_{c}$.
To analyze the shift of these sideband peaks caused by the force $f$, we compare the shifted emission spectrum with the original emission spectrum which is obtained in the absence of force $f$. When the force $f$ is applied to the moving mirror, the resonance sideband peaks in the emission spectrum will be shifted in the frequency space by $2g_{0}\eta/\omega_{M}$ which is a linear function of the force magnitude $f$ because of $\eta=fx_{0}$. To characterize this frequency shift, it is needed to choose a peak as the reference. In this system, however, there are many phonon sideband peaks in the spectrum. These phonon sideband peaks might cause confusion when we decide the frequency shift induced by the external force, because the shifted peak might stride several phonon sideband peaks. Therefore, a detailed analysis on the dependence of the peak shift on the force magnitude is needed to evaluate the force. Concretely, we consider two cases: $\eta>0$ and $\eta<0$, which correspond to the situations in which the force and the defined positive mechanical oscillation are in the same and opposite directions, respectively. When $\eta>0$, we have $-(g_{0}^{2}+2g_{0}\eta)/\omega_{M}<0$, which means that the ZPL is in the red-sideband zone. Depending on the relation between the parameters $g_{0}$, $\eta$, and $\omega_{M}$, the ZPL could be in the range $-(m+1)\omega_{M}<-(g_{0}^{2}+2g_{0}\eta)/\omega_{M}<-m\omega_{M}$ for a natural number $m$, and then the first (labeled from right to left in the region $\Delta_{k} < 0$) peak in the red sideband is located at $\Delta_{k}=m\omega_{M}-(g_{0}^{2}+2g_{0}\eta)/\omega_{M}$. For example, when $-\omega_{M}<-(g_{0}^{2}+2g_{0}\eta)/\omega_{M}<0$, then the first peak in the red sideband appears at $\Delta_{k}=-(g_{0}^{2}+2g_{0}\eta)/\omega_{M}$. When $\eta<0$, the ZPL shift $-(g_{0}^{2}+2g_{0}\eta )/\omega_{M}$ could be either positive or negative. Similar to the positive $\eta$ case, if $-(m+1)\omega_{M}<-(g_{0}^{2}+2g_{0}\eta )/\omega_{M}<-m\omega_{M}$ for a natural number $m$, then the first peak in the red sideband is located at $\Delta_{k}=m\omega_{M}-(g_{0}^{2}+2g_{0}\eta )/\omega_{M}$. When the ZPL satisfies $m\omega_{M}<-(g_{0}^{2}+2g_{0}\eta )/\omega_{M}<(m+1)\omega_{M}$, then the first peak in the red sideband appears at $\Delta_{k}=-(g_{0}^{2}+2g_{0}\eta)/\omega_{M}-(m+1)\omega_{M}$.
\begin{figure}[tbp]
\center
\includegraphics[width=9cm]{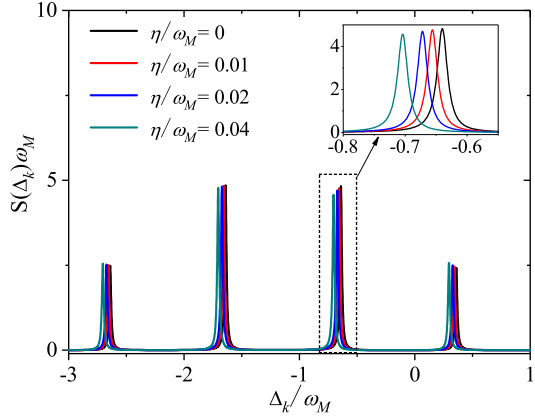}
\caption{The scaled emission spectrum $S(\Delta_{k})\omega_{M}$ as a function of $\Delta_{k}/\omega_{M}$ when the force parameter takes various values $\eta/\omega_{M}=0$, $0.01$, $0.02$, and $0.04$. The initial state of the moving mirror is $|0\rangle_{b}$, and other parameters are the same as those in Fig.~\ref{Fig2}.}
\label{Fig3}
\end{figure}

In Fig.~\ref{Fig3}, we show the single-photon emission spectrum of the optomechanical cavity when the magnitude of the external force takes various values: $\eta/\omega_{M}=0.01$, $0.02$ and $0.04$. We also present the emission spectrum in the absence of the force (i.e., $\eta=0$) as a reference. Here the initial state of the movable mirror is taken as $|0\rangle_{b}$. From Fig.~\ref{Fig3} we can see that the resonant peaks in the emission spectrum will be shifted due to the action of the external force. By checking the parameters and the peak shifts, we find that the locations of these sideband peaks are shifted by $2g_{0}\eta/\omega_{M}$, as expected by the above analyses. Since these peak shifts corresponding to the used values do not cross the sidebands, the force magnitude can be inferred directly from the spectrum shift. For the parameters used in Fig.~\ref{Fig3}, the force parameter should satisfy the condition $\eta>\gamma\omega_{M}/(2g_{0})=0.0125\omega_{M}$ to resolve the peak shift from the peak width, in the sense that the distance between the shifted peak and the peak of reference should be larger than the full width at half maximum of the peaks. Here we can see that the peak shift in the case of $\eta/\omega_{M}=0.01$ can not be resolved from the original peak, while it can be resolved in the cases of $\eta/\omega_{M}=0.02$ and $0.04$ (see the inset). We point out that the undetectable bath of the optomechanical cavity affects the width of the sideband peaks, which in turn determines the smallest measurable force through the resolution of the shift of these peaks.

For a sufficient large force, the peak shift might cross several sidebands, then how to decide the peak of reference becomes a critical problem. A possible method for solving this problem is a pre-estimation of the magnitude of the force. By analyzing the correct frequency shift of the sideband peaks, then the magnitude of the force can be inferred. Otherwise, if a wrong peak is selected, then the value of the estimated external force will differ the correct force by the error $\Delta f=l\omega_{M}^{2}/(2g_{0}x_{0})$ with $l$ being the number of the miscounted sidebands. Here we assume that the error of the frequency shift is only caused by a wrong count of the number of crossed sidebands. Below, we give some evaluation on the detection sensitivity of the force. In this system, the width of these sideband peaks is approximately $\gamma$, then the measurement accuracy of the $\eta$ is $\gamma\omega_{M}/(2g_{0})$ which is determined by the condition $2g_{0}\eta/\omega_{M}>\gamma$. In the resolved-sideband regime $\gamma\ll\omega_{M}$, the measurement error $\Delta\eta=\Delta fx_{0}=l\omega_{M}^{2}/(2g_{0})$ is much greater than the measurement accuracy, which means that the main reason for inaccurate measurement is the wrong estimation of the peak shift.
\begin{figure}[tbp]
\center
\includegraphics[width=8cm]{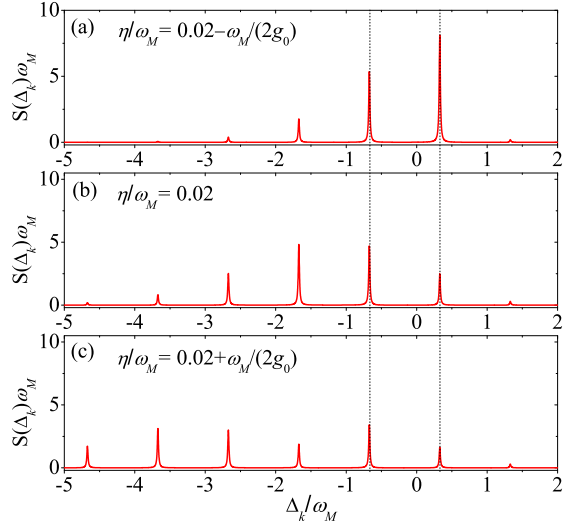}
\caption{The scaled emission spectrum $S(\Delta_{k})\omega_{M}$ as a function of $\Delta_{k}/\omega_{M}$ when the force parameter takes $\eta/\omega_{M}=0.02$ and $0.02\pm\omega_{M}/(2g_{0})$. The peak shifts in these three cases differ from each other by $\pm\omega_{M}$. The initial state of the moving mirror is $|0\rangle_{b}$, and other parameters are the same as those in Fig.~\ref{Fig2}.}
\label{Fig4}
\end{figure}

To see the phenomenon of the peak shifts crossing sidebands, in Fig.~\ref{Fig4} we display the single-photon emission spectrum when the force parameter takes different values $\eta/\omega_{M}=0.02$ and $0.02\pm\omega_{M}/(2g_{0})$. Here the amplitudes $\omega_{M}/(2g_{0})$ and $-\omega_{M}/(2g_{0})$ correspond to the peak shift crossing right one sideband peak towards the left and right, respectively. Figure~\ref{Fig4} shows that the locations of these sideband peaks in these emission spectra are the same. However, the spectra for $\eta/\omega_{M}=0.02\pm\omega_{M}/(2g_{0})$ can not obtained by a direct displacement of the spectrum for $\eta/\omega_{M}=0.02$ in the frequency space. This is because the values of these Franck-Condon factors in Eqs.~(\ref{SOLlongt}) are also determined by the value of $\eta$.

The correctness of the choice of the peak shift can be checked as follows. We use the obtained value of $\eta$ to calculate a theoretic emission spectrum, and then compare this theoretic emission spectrum with the spectrum obtained in experiments. If the theoretic spectrum matches well with the experimental spectrum, then the selected peak shift is the true value, and the magnitude of the external force can be inferred correctly from the value of $\eta$. Otherwise, other peaks need to be tried until the correct one is obtained. In a realistic experiment, we focus on the ZPL determined by $\Delta_{k}=-(g_{0}^{2}+2g_{0}\eta )/\omega_{M}$. If the ZPL is located at $\Delta_{k0}$ then the magnitude of the external force can be inferred as $f=-(\Delta_{k0}\omega_{M}+g_{0}^{2})/(2x_{0}g_{0})$.

\subsection{Force detection by scattering spectrum}

The force detection scheme can also be implemented by analyzing single-photon scattering spectrum. In the scattering case, the single photon is initially in the detected continuous fields outside the cavity. In particular, we consider the case where the single photon is initially in the Lorentzian wavepacket, and there are no photons in the cavity and the undetected vacuum bath. Similar to the emission case, we first solve the scattering problem for the case where the mirror is in a number state $|m_{0}\rangle_{b}$. Once the solution for this case is obtained, the solution corresponding to an arbitrary initial state of the mirror can be obtained using the same procedure as the emission case.
Below, we consider the initial state of the system as $|\varphi (0)\rangle =\sqrt{\epsilon /[\pi \mu (\omega_{c})]}\sum_{k}(\Delta _{k}-\Delta _{0}+i\epsilon )^{-1}|0\rangle _{a}|m_{0}\rangle _{b}|1_{k}\rangle _{c}|\emptyset \rangle _{d}$ where $\Delta_{0}$ and $\epsilon$ are the detuning and spectral width of the photon, respectively. The initial conditions of these probability amplitudes corresponding to the initial state are give by $A_{m}(0)=C_{m,q}(0)=0$ and$\hspace{0.2cm}B_{m,k}(0)=\sqrt{\epsilon /[\pi \mu (\omega _{c})]}(\Delta _{k}-\Delta _{0}+i\epsilon
)^{-1}\,_{b}\!\langle \tilde{m}(0)|m_{0}\rangle _{b}$.
Based on the initial conditions, the transient solution of these probability amplitudes can be obtained. For calculation of the scattering spectrum, we focus on the long-time solution under the condition $(\max \{1/\gamma,1/2\epsilon\} \ll t\ll 1/\gamma_{M})$. Similar to the emission case, we add an index $m_{0}$ to these probability amplitudes to mark the initial state $|m_{0}\rangle_{b}$ of the moving mirror. The expressions of the single-photon scattering solution are
\begin{subequations}
\label{scattsolut}
\begin{align}
A_{m_{0},m}(t\rightarrow\infty )& =0, \\
B_{m_{0},m,k}(t\rightarrow\infty )& =\sum_{l=0}^{\infty}\,_{b}\langle \tilde{l}(0)|m_{0}\rangle_{b}B_{m,k}^{l}(t\rightarrow\infty), \\
C_{m_{0},m,q}(t\rightarrow\infty )& =\sum_{l=0}^{\infty}\,_{b}\langle \tilde{l}(0)|m_{0}\rangle_{b}C_{m,q}^{l}(t\rightarrow\infty),
\end{align}
\end{subequations}
where $B_{m,k}^{l}(t\rightarrow\infty)$ and $C_{m,q}^{l}(t\rightarrow\infty)$ are, respectively, the long-time solutions of the probability amplitudes corresponding to the initial state $|\tilde{l}(0)\rangle_{b}$ of the moving mirror. This is because the initial state $|m_{0}\rangle_{b}$ can be written as $\sum_{l=0}^{\infty}\,_{b}\langle \tilde{l}(0)|m_{0}\rangle_{b}|\tilde{l}(0)\rangle_{b}$. The expressions of $B_{m,k}^{l}(t\rightarrow\infty)$ and $C_{m,q}^{l}(t\rightarrow\infty)$ are given by
\begin{subequations}
\label{the solutions}
\begin{align}
B_{m,k}^{l}(t\rightarrow \infty )=& \sqrt{\frac{\epsilon }{\pi \mu (\omega
_{c})}}\frac{\delta _{m,l}e^{-i(\Delta _{k}+m\omega _{M}-\zeta )t}}{\Delta
_{k}-\Delta _{0}+i\epsilon }-\sum_{n=0}^{\infty }\frac{\,_{b}\langle \tilde{m%
}(0)|\tilde{n}(1)\rangle _{b}\,_{b}\langle \tilde{n}(1)|\tilde{l}(0)\rangle
_{b}}{\Delta _{k}+\lambda -(n-m)\omega _{M}+i\frac{\gamma }{2}}  \notag \\
& \times \frac{i\gamma _{c}\sqrt{\frac{\epsilon }{\pi \mu (\omega _{c})}}%
e^{-i(\Delta _{k}+m\omega _{M}-\zeta )t}}{\Delta _{k}-\Delta
_{0}-(l-m)\omega _{M}+i\epsilon }, \\
C_{m,q}^{l}(t\rightarrow \infty )=& -\sum_{n=0}^{\infty }\frac{\,_{b}\langle
\tilde{m}(0)|\tilde{n}(1)\rangle _{b}\,_{b}\langle \tilde{n}(1)|\tilde{l}%
(0)\rangle _{b}}{\Delta _{q}+\lambda -(n-m)\omega _{M}+i\frac{\gamma }{2}}%
\frac{i\sqrt{\gamma _{c}\gamma _{d}}\sqrt{\frac{\epsilon }{\pi \nu (\omega
_{c})}}e^{-i(\Delta _{q}+m\omega _{M}-\zeta )t}}{\Delta _{q}-\Delta
_{0}-(l-m)\omega _{M}+i\epsilon }.
\end{align}
\end{subequations}
The solution $A_{m_{0},m}(t\rightarrow\infty)=0$ in Eq.~(\ref{scattsolut}) implies that the single photon leaks completely out of the cavity in the long-time limit. The first term of $B_{m,k}^{l}(t\rightarrow\infty)$ represents a physical process in which the single injected photon is directly reflected by a mirror without entering the cavity, and thus there is no energy level transition in this term. The second term of $B_{m,k}^{l}(t\rightarrow\infty )$ describes the physical process in which the single photon enters the cavity and then emits out of the cavity.

The above obtained scattering solution corresponds to the initial number state of the movable mirror. In principle, when the movable mirror is initially in an arbitrary state, the scattering spectrum can also be obtained. Corresponding to either the pure state $\sum_{m_{0}=0}^{\infty}R_{m_{0}}(0)|m_{0}\rangle_{b}$ or the mixed state $\sum_{m_{0}=0}^{\infty}P_{m_{0}}(0)|m_{0}\rangle_{b}\!\,_{b}\langle m_{0}|$, the spectra can be obtained by the formula in Eqs.~(\ref{purespectra}) and~(\ref{mixedspectra}). It should be noted that the form of the wave packet of injection will affect the waveform of the scattering spectrum. For the injected Lorentzian wavepacket, the width of the wave packet will determine the scattering spectrum. Below we consider two different cases of single-photon scattering: the wide and narrow wavepacket injections.
\begin{figure}[tbp]
\center
\includegraphics[width=11cm]{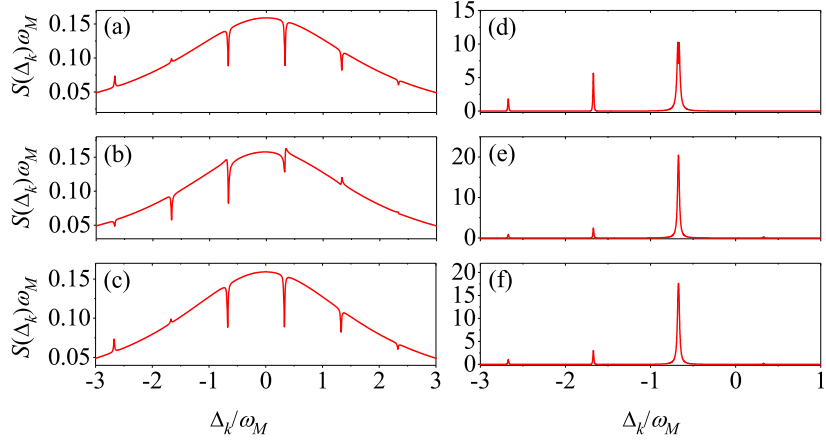}
\caption{The left column is the scaled scattering spectrum in the wide-wavepacket-injection case, when the moving mirror is initially prepared in various states: (a) ground state $|0\rangle_{b}$, (b) coherent state $|\alpha\rangle_{b}$ with $\alpha=1$, and (c) thermal state $\rho_{b}^{\text{th}}$ with $\bar{n}_{\text{th}}=1$. The used parameters are $g_{0}/\omega_{M}=0.8$, $\gamma_{c}/\omega_{M}=\gamma_{d}/\omega_{M}=0.01$,  $\eta/\omega_{M}=0.02$, $\epsilon/\omega_{M}=2$, and $\Delta_{0}/\omega_{M}=0$. The right column is the scaled scattering spectrum in the narrow-wavepacket-injection case, when the moving mirror is in (d) ground state $|0\rangle_{b}$, (e) coherent state $|\alpha\rangle_{b}$ with $\alpha=1$, and (f) thermal state $\rho_{b}^{\text{th}}$ with $\bar{n}_{\text{th}}=1$. The used parameters are $g_{0}/\omega_{M}=0.8$, $\gamma_{c}/\omega_{M}=\gamma_{d}/\omega_{M}=0.01$,  $\eta/\omega_{M}=0.02$, $\epsilon/\omega_{M}=0.01$, and $\Delta_{0}=-\lambda$.}
\label{Fig5}
\end{figure}

In the left column of Fig.~\ref{Fig5}, we plot the single-photon scattering spectrum when the mirror is initially in different states: ground state, coherent state, and thermal state. Here we consider the wide-wavepacket-injection case by choosing $\epsilon/\omega_{M}=2$. When the width $\epsilon$ of the injected wavepacket is larger than the mechanical frequency $\omega_{M}$, there always exist resonant transition channels and the photon-mirror interaction process (described by the the second term in $B_{m,k}^{l}(t\rightarrow\infty)$) becomes important. The the left column of Fig.~\ref{Fig5} shows that, in the single-photon scattering spectrum, there are both minor peaks and dips attached on the main Lorentzian spectrum, which are formed by quantum interference effect between the direct reflected photon and the scattered photon. For different initial states, the locations of these peaks and dips are the same, because these peaks and dips are determined by the system parameters and correspond to the transitions between the energy levels in the zero-photon and one-photon wells [Fig.~\ref{Fig1}(b)]. As a result, the shift of these peaks can be used to infer the magnitude of the external force. However, the height of these peaks and the depth of these dips are different because the magnitudes of these transitions depend on the initial state of the moving mirror.

We also plot the scattering spectrum corresponding to various mirror's initial states in the narrow-wavepacket-injection case (see the right column of Fig.~\ref{Fig5}). Here we choose the wavepacket center of the injected photon as $\Delta_{0}=-\lambda$ so that the injected photon can resonantly enter the cavity and the photon injection will induce resonant transition $|0\rangle_{a}|\tilde{0}(0)\rangle_{b}\rightarrow|1\rangle_{a}|\tilde{0}(1)\rangle_{b}$. The width of the injected wavepacket is chosen as $\epsilon/\omega_{M}=0.01$ such that it matches the linewidth of the detected cavity emission channel. The right column of Fig.~\ref{Fig5} shows that, in the narrow-wavepacket case, there mainly exist peaks in the spectrum and the locations of these peaks are independent of the initial states of the moving mirror. Since the injected photon can resonantly enter the cavity, then the reflected wavepacket and the scattered wavepacket will be overlapped in the frequency space. This is because the scattered photon is mostly determined by the resonant transition $|1\rangle_{a}|\tilde{0}(1)\rangle_{b}\rightarrow|0\rangle_{a}|\tilde{0}(0)\rangle_{b}$ and hence the locations of these sideband peaks are determined by the eigenenergy spectrum of the system. As a result, the main peak in the scattering spectrum will be located at the $\Delta_{k}=-\lambda$, as shown in the right column of Fig.~\ref{Fig5}.

\begin{figure}[tbp]
\center
\includegraphics[width=11cm]{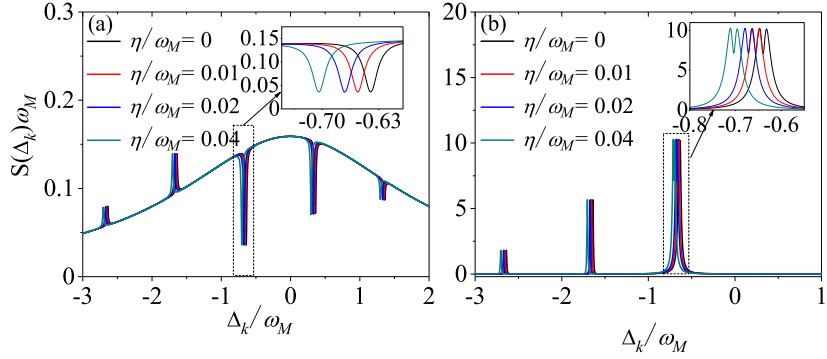}
\caption{(a) The scaled scattering spectrum $S(\Delta_{k})\omega_{M}$ as a function of $\Delta_{k}/\omega_{M}$ in the wide-wavepacket-injection case. The force parameter takes various various values $\eta/\omega_{M}=0$, $0.01$, $0.02$, and $0.04$. The initial state of the moving mirror is $|0\rangle_{b}$, other parameters are the same as those in Fig.~\ref{Fig5}(a). (b) The scaled scattering spectrum $S(\Delta_{k})\omega_{M}$ as a function of $\Delta_{k}/\omega_{M}$ in the narrow-wavepacket-injection case. The force parameter takes various values $\eta/\omega_{M}=0$, $0.01$, $0.02$, and $0.04$. The initial state of the moving mirror is $|0\rangle_{b}$, and other parameters are the same as those in Fig.~\ref{Fig5}(d).}
\label{Fig6}
\end{figure}
The scattering spectrum has a similar dependence relation on the external force and hence it can also be used to detect the external force. To show the dependence of either the peak shift or the dip shift on the force magnitude, in Fig.~\ref{Fig6}(a) we plot the single-photon scattering spectrum as a function of $\Delta_{k}$ when the force parameter takes different values: $\eta/\omega_{M}=0.01$, $0.02$, and $0.04$. Here we can see that the shifts of these peaks and dips depend on the magnitude of the force. Similar to the emission case, the shift of these peaks and dips can be resolved from the original peak- and dip-width when the force magnitude satisfies the relation $\eta>\omega_{M}\gamma/(2g_{0})=0.0125\omega_{M}$. As shown in the inset, only the two peak shifts corresponding to $\eta/\omega_{M}=0.02$ and $0.04$ can be resolved from the cavity field linewidth.

We also investigated the dependence of the peak shift on the external force in the narrow-wavepacket scattering case. The single-photon scattering spectrum in this case is plotted in Fig.~\ref{Fig6}(b). Here we can see that the peak shift is proportional to the force magnitude, and then the force magnitude can be inferred from the peak shift. In particular, we can see a dip in the main peak (see the inset), this dip is caused by quantum interference between the directly reflected photon and the scattered photon. In Fig.~\ref{Fig6}(b), we choose $\Delta_{0}=-\lambda$, which corresponds to the resonant transition $|0\rangle_{a}|\tilde{0}(0)\rangle_{b}\rightarrow|1\rangle_{a}|\tilde{0}(1)\rangle_{b}$. The scattered photon associated with the inverse process of this transition has the same wave-packet center with the reflected photon, and then the quantum interference effect induce observable dip in the main peak. It should be noted that the physical mechanism of the force detection works in both the resonant and nonresonant injection cases. Here we only exhibit the resonant injection case because the entering probability for the single photon is enhanced when it induces the resonant transitions. In the nonresonant injection case, the force detection can be realized by analyzing the scattering peaks corresponding to the photon entering the cavity. These scattering peaks are weak because the photon is mainly reflected by the fixed mirror (without entering the cavity).

\section{Detection of a weak periodic force \label{sec4}}

The above discussions are restricted to the detection of a weak constant force. In this section, we study how to detect a monochromatic oscillating weak force $f(t)=f\cos(\omega_{f}t) $ acting on the movable mirror, where $\omega_{f}$ is the resonance frequency of the force. For utilizing the spectrometric method to detect the monochromatic oscillating weak force, we need to know the oscillating frequency $\omega_{f}$ of the periodic force in advance, and then we introduce a periodic modulation with the same frequency $\omega_{f}$ as the oscillating force to the optomechanical coupling strength: $g(t)=g_{0}\cos(\omega_{f}t)$~\cite{Liao2014NJP}.
Using the same notation as those in the above sections, the Hamiltonian of the total system including the optomechanical cavity and its vacuum bath can be expressed as $\hat{H}^{\prime }=\hat{H}_{\text{opc}}^{^{\prime }}+\hat{H}_{\text{B}}+\hat{H}_{\text{int}}$
where the optomechanical Hamiltonian in the modulated case reads
\begin{equation}
\hat{H}_{\text{opc}}^{^{\prime}}=\omega_{c}\hat{a}^{\dagger}\hat{a}+\omega_{M}\hat{b}^{\dagger}\hat{b}-(g_{0}\hat{a}^{\dagger}\hat{a}+\eta)(\hat{b}^{\dagger}+\hat{b})\cos (\omega_{f}t).
\end{equation}

In a rotating frame with respect to $\hat{H}_{0}^{\prime }=\omega _{c}(\hat{a}^{\dagger }\hat{a}+\sum_{k}%
\hat{c}_{k}^{\dagger }\hat{c}_{k}+\sum_{q}\hat{d}_{q}^{\dagger }\hat{d}_{q})+\omega _{f}\hat{b}^{\dagger }\hat{b}$, we can obtain a Hamiltonian of the total system in the rotating frame. For force detection, we consider the parameter conditions $\omega_{f}\gg 0$ and $\omega_{M}+\omega_{f}\gg (g_{0}\langle\hat{a}^{\dagger}\hat{a}\rangle+\eta)/2$, and then by making the rotating-wave approximation we obtain
\begin{eqnarray}
\hat{H}_{I}^{\prime } &=&\omega _{M}^{\prime }\hat{b}^{\dagger }\hat{b}%
-(g_{0}^{\prime }\hat{a}^{\dagger }\hat{a}+\eta ^{\prime })(\hat{b}^{\dagger
}+\hat{b})+\sum_{k}\Delta _{k}\hat{c}_{k}^{\dagger }\hat{c}%
_{k}+\sum_{q}\Delta _{q}\hat{d}_{q}^{\dagger }\hat{d}_{q}  \notag \\
&&+\sum_{k}\xi _{k}(\hat{c}_{k}^{\dagger }\hat{a}+\hat{a}^{\dagger }\hat{c}%
_{k})+\sum_{q}\chi _{q}(\hat{a}^{\dagger }\hat{d}_{q}+\hat{d}_{q}^{\dagger }%
\hat{a}),
\end{eqnarray}
where introduce the variables $\omega'_{M}=\omega_{M}-\omega_{f}$, $g'_{0}=g_{0}/2$, and $\eta' =\eta/2$. This Hamiltonian has the same form as that in Eq.~(\ref{eq1iniH}) when $\omega_{M}>\omega_{f}$, so the studies on the constant force detection can be used to detect the oscillating force.

\section{Discussions and conclusion \label{sec5}}

We now present some discussions on the parameter requirement of this scheme and the precision of the force detection. In this scheme, the optomechanical cavity should work in the resolved-sideband regime $\gamma/\omega_{M}\ll 1$ and the single-photon strong-coupling regime $g_{0}>\gamma$ so that the sideband peaks can be used to characterize the force detection. This requires that the decay rate of the cavity should be as small as possible. To resolve the peak shift from the cavity line width, the condition $2g_{0}\eta/\omega_{M}\geq\gamma$ should be satisfied. Then the magnitude of the minimum measurable force can be obtained as $\hbar \gamma \omega_{M}/(2g_{0}x_{0})$. According to current experimental conditions~\cite{Aspelmeyer2014RMP}, we take a set of parameters corresponding to the single-photon strong-coupling regime: $\gamma /\omega_{M}\approx0.01$, $g_{0}/\omega_{M}\approx1$, $\omega_{M}\approx2\pi\times100$ MHz, and the zero-point fluctuation $x_{0}\sim0.4\times10^{-14}$ m, then the minimum measurable force magnitude can reach $f\approx8.25\times10^{-14}$ N.

It should be pointed out that a smaller force can also be measured by analyzing the change of the height of the single-photon spectrum rather than the shift of the resonance peaks in the spectrum. For example, we choose a point (for example $\Delta_{k'}$) in the spectrum corresponding to the absence of the force as a reference, and the height of the spectrum at this point can be obtained. When the force is applied upon the resonator, we get another value of the height of the spectrum at the same point $\Delta_{k'}$, then the difference between the two heights of the spectrum at this sample point is caused by the force. Therefore, by analyzing the change of the spectral height at this point, the magnitude of the force can be inferred by comparing the detected spectrum with the theoretically calculated spectrum. In principle, the minimum measurable force magnitude with this method could be very small because it is determined by the sensitivity of the observed spectral difference and the precision of the theoretical deduction of the force magnitude from the spectrum change. Here, we should mention that the magnitude of the force to be detected should keep consistent with the validity of the present physical model. Namely the force to be detected is described by a classical force. In the classical force case, the force is a parameter and we do not need to consider the back action induced by the detection. However, the back action should be considered in the quantum force case because the force is a dynamical variable.

In conclusion, we proposed a spectroscopic method to detect a classical force, which is acted on the moving end mirror of an optomechanical cavity. We obtained the analytical solutions of the single photon emission and scattering problem and the corresponding spectra. We presented the parameter conditions under which the magnitude of the force can be inferred from the spectra. We also extended this spectrometric method to detect the magnitude of a monochromatic oscillating force by introducing a time modulation to the optomechanical coupling strength.

\section*{Funding}
Q.S.T. is supported in part by the NSFC Grants No. 11665010 and No. 11805047. J.-F.H. is supported in part by the National Natural Science Foundation of China (Grant No.~11505055) and Scientific Research Fund of Hunan Provincial Education Department (Grant No. 18A007). J.-Q.L. is supported in part by National Natural Science Foundation of China (Grants No.~11822501, No.~11774087, and No.~11935006), Natural Science Foundation of Hunan Province, China (Grant No.~2017JJ1021), and Hunan Science and Technology Plan Project (Grant No.~2017XK2018).

\section*{Disclosures}
The authors declare no conflicts of interest.


\end{document}